\documentclass{nature}
\usepackage{graphicx}
\usepackage{amsmath}
\usepackage{color}
\usepackage{physics}
\usepackage{hyperref}
\usepackage{color}
\usepackage{soul}
\usepackage{slashed}
\usepackage[labelformat=empty]{caption}

\newcommand{\blue}[1]{{\color{blue} #1}}

\makeatletter
\let\saved@includegraphics\includegraphics

\title{\blue{An Idealised Approach of Geometry and Topology to the Diffusion of Cations in Honeycomb Layered Oxide Frameworks}}

\author{Godwill Mbiti Kanyolo$^1$ \& Titus Masese$^2$$^{,3}$}

\begin{document}

\maketitle

\begin{affiliations}
 \item Department of Engineering Science, The University of Electro-Communications 1-5-1 Chofugaoka, Chofu, Tokyo 182-8585, Japan
 
 \item Research Institute of Electrochemical Energy (RIECEN), National Institute of Advanced Industrial Science and Technology (AIST), 1-8-31 Midorigaoka, Ikeda, Osaka, 563-8577, Japan
 
 \item AIST-Kyoto University Chemical Energy Materials Open Innovation Laboratory (ChEM-OIL), Sakyo-ku, Kyoto 606-8501, Japan
 
\end{affiliations}

\begin{abstract}
Honeycomb layered oxides are a novel class of nanostructured materials comprising alkali or alkaline earth metals intercalated into transition metal slabs. The intricate honeycomb architecture and layered framework endows this family of oxides with a tessellation of features such as exquisite electrochemistry, unique topology and fascinating electromagnetic phenomena. Despite having innumerable functionalities, these materials remain highly underutilized as their underlying atomistic mechanisms are vastly unexplored. Therefore, in a bid to provide a more in-depth perspective, we propose an idealised diffusion model of the charged alkali cations (such as lithium, sodium or potassium) in the two-dimensional (2D) honeycomb layers within the three-dimensional (3D) crystal of honeycomb layered oxide frameworks. This model not only explains the correlation between the excitation of cationic vacancies (by applied electromagnetic fields) and the Gaussian curvature deformation of the 2D surface, but also takes into consideration, the quantum properties of the cations and their inter-layer mixing through quantum tunnelling. Through this work, we offer a novel theoretical framework for the study of 3D layered materials with 2D cationic diffusion currents, as well as providing pedagogical insights into the role of topological phase transitions in these materials in relation to Brownian motion and quantum geometry.
\end{abstract}

\begin{figure}[!b]
\begin{center}\label{Rendition}
\includegraphics[width=1\columnwidth,clip=true]{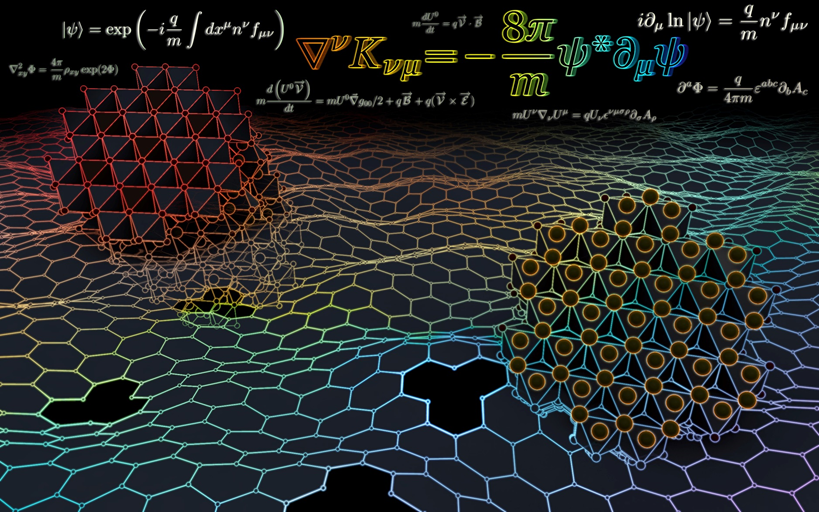}
  \caption{\textbf{Figure 1: A rendition displaying the honeycomb layer of alkali cations in honeycomb layered oxides as curved stacked two-dimensional (2D) manifolds forming a three-dimensional (3D) crystal.} \blue{The main equation from our model links curvature variations to energy-momentum within the material responsible for diffusion currents and topological transitions.}}
\end{center}
\end{figure}

\newpage

\section{Introduction}

Nanotechnology has become the cornerstone of contemporary science for its role in the discovery of new materials with unprecedented chemical properties and unconventional physical phenomena. Typically these stellar properties are optimised and refined through manipulation of matter at an atomic or molecular level. As such, the fundamental understanding of the physical laws surrounding the interaction of atoms and atom clusters in the different phases of matter is invaluable in the evolution of this technology. Theoretical advancements into the connection between continuous symmetry and conservation laws (\textbf{Noether's theorem\footnote{For every continuous symmetry of the action, there is a corresponding conservation law.}}) in quantum theory and geometry has played an enormous role in revealing exemplary quantum effects in condensed matter systems. This has singled out materials comprising simple geometric arrangements of atoms as exemplars of Noether's theorem while offering a great segue into crystallography. This rationale has generated interest particularly in two dimensional (2D) honeycomb layered oxides whose heterostructural layout plays host to an assortment of desirable electrochemical, magnetic and topological properties.\cite{Masese2018, Sathiya2013, Yang2017, Yuan2014, Gupta2013, Masese2019, Masese2019b, Evstigneeva2011, Zheng2016, Ma2015, Nalbandyan2013,Kumar2012, Zvereva2013, Derakhshan2007, Viciu2007, Morimoto2007, Derakhsan2008, Zvereva2012, Upadhyay2016, Koo2016, Zvereva2016, Zvereva2017, Kumar2013, Berthelot2012, Laha2013, Taylor2019, Roudebush2013, Uma2016, He2017, He2018, Schmidt2013, Heymann2017, Bette2019, Bhardwaj2014}

Honeycomb layered oxides generally adopt the following compositions:\\
${\rm A^{+}_{3}L^{3+}D^{4+}O_{5} (A^{+}_{6/5}L_{2/5}^{3+}D_{2/5}^{4+}O_{2}), A^{+}_{4}L^{2+}D^{6+}O_{6}\, (A^{+}_{8}L^{2+}_{2}D^{6+}_{2}O_{12}}$ or ${\rm A^{+}_{4/3}L^{2+}_{1/3}D^{6+}_{1/3}O_{2}),}$\\ 
${\rm A^{+}_{2}L^{3+}D^{5+}O_{5}\, (A^{+}_{4/5}L_{2/5}^{3+}D_{2/5}^{5+}O_{2}), A^{+}_{3}L^{2+}_{2}D^{5+}O_{6} (A^{+}L^{2+}_{2/3}D^{5+}_{1/3}O_{2}),A^{+}_{2}L^{2+}D^{4+}O_{6} \,(A^{+}_{2/3}L_{1/2}^{2+}D_{1/3}^{4+}O_{2}),}$\\
${\rm A^{+}_{3}A^{*+}L^{3+}D^{5+}O_{6}}$ ${\rm (A^{+}A^{*+}_{1/3}L^{3+}_{1/3}D^{5+}_{1/3}O_{2}),A^{+}_{2}L^{2+}_{2}D^{6+}O_{6} (A^{+}_{2/3}L^{2+}_{2/3}D^{6+}_{1/3}O_{2}), A^{+}_{2}L_{3}^{2+}D^{4+}O_{6}}$ \\
(or equivalently as ${\rm A^{+}_{2/3}L^{2+}D_{1/3}^{4+}O_{2}), A^{+}_{4}L^{3+}D^{5+}O_{6} (A^{+}_{8}L^{3+}_{2}D^{5+}_{2}O_{12}}$ or ${\rm A^{+}_{4/3}L^{3+}_{1/3}D^{5+}_{1/3}O_{2}),}$\\
${\rm A^{+}_{2}D^{4+}O_{3} (A^{+}_{4/3}D^{4+}_{2/3}O_{2}), A^{+}_{4.5}L^{3+}_{0.5}D^{6+}O_{6}}$ ${\rm(A^{+}_{1/2}L^{3+}_{1/6}D^{6+}_{1/3}O_{2}),}$
where ${\rm L}$ can be ${\rm Zn, Mn, Fe, Co, Cu,}\\ 
{\rm Ni, Cr, Mg}$; ${\rm D}$ can be ${\rm Bi, Te, Sb, Ta, Ir, Nb, W, Sn, Ru, Mo}$; ${\rm A}$ and ${\rm A^{*}}$ can be alkali atoms (such as ${\rm Li, Cu, K, Rb, Cs, Ag}$ and ${\rm Na}$ with ${\rm A \neq A^{*}}$), transition metal atoms, for instance ${\rm Cu}$ or noble metal atoms (e.g., ${\rm Ag, Au, Pd,}$ etc).

From a crystal outlook, this family of layered oxides consists mainly of alkali cations (labeled in the above list of compositions as ${\rm A}^{+}$) sandwiched between parallel slabs (stackings) of transition metal oxides (${\rm L{\rm O}_{6}}$ and ${\rm D{\rm O}_{6}}$ octahedra). Some of the oxygen (${\rm O}$) atoms coordinate with ${\rm A^{+}}$ cations to form inter-layer bonds whose strength is dependent on the inter-layer distance between the slabs. In fact, there is a correlation between the stacking structure and the resulting electrochemical performance of the honeycomb layered oxides that can be traced to the differing sizes of the ${\rm A^{+}}$ cations. For instance, ${\rm A^{+}}$ cations with small ionic radii such as ${\rm Li^{+}}$ tend to form stronger inter-layer bonds as a result of the smaller inter-layer distance. However, ${\rm K^{+}}$ has a vastly larger ionic radius with a correspondingly larger inter-layer distance and hence forms weaker inter-layer bonds. Generally, ${\rm A^{+}}$  cations with larger ionic radii such as ${\rm K^{+}}$ and ${\rm Na^{+}}$ form weaker interlayer bonds in the aforementioned compositions resulting in layered oxides with prismatic or octahedral coordination of alkali metal and oxygen (technically referred to as P-type or O-type layered structures, respectively). The weaker interlayer bonds in prismatic layered (P-type) structures create more open voids within the transition metal layers allowing for facile two-dimensional diffusion of alkali atoms within the slabs.  This gives rise to the high ionic mobility and exceptional electrochemical properties innate in honeycomb layered oxides.

In our study, we focus on the prismatic subclass of honeycomb layered oxides that generally adopt ${\rm A^{+}_{2}L_{2}^{2+}D^{6+}O_{6}}$ (or equivalently ${\rm A^{+}_{2/3}L_{2/3}^{2+}D_{1/3}^{6+}O_{2}}$) compositions, where ${\rm A = K, Li}$ or ${\rm Na}$ is an alkali cation (potassium, lithium or sodium) owing to their exemplary electrochemical and physical properties. We explore their cationic diffusion by envisioning an idealised model of three-dimensional (3D) layered oxides in an attempt to gain an effective description of the diffusion mechanics along the honeycomb layers using concepts of 2D curvature and topology. We proceed to link geometric properties such as the Gaussian curvature and the genus of the 2D honeycomb layers to transport quantities of the cations such as their charge density and cationic vacancies respectively. The inter-layers can act as tunnel barriers quantum mechanically traversable by the cations. The proposed model is solved by identifying symmetries in a curved space-time implemented by Killing vectors along the time ($t$) and longitudinal ($z$) directions.In order to discern the connections between the honeycomb topology, applied magnetic fields and other crystalline symmetries, a multidimensional approach integrating techniques and concepts from various fields were employed. As such, the results presented herein bear particular significance across a diversity of disciplines ranging from topological order and phase transitions in materials to their relation to Brownian motion and quantum geometries.

Throughout the paper, we set Planck's constant and the speed of light in the crystal to unity ($\hbar = \bar{c} = 1$.) We employ Einstein's summation convention together with the Minkwoski 2D + 1 and 3D + 1 metrics, $g_{ab} \equiv {\rm diag}(1, -1, -1)$ and $g_{\mu\nu} \equiv {\rm diag}(1, -1, -1, -1)$ to lower the Roman and Greek indices respectively. The Roman indices $i,j,k$ are reserved for Euclidean space $g_{ij} \equiv {\rm diag}(1, 1, 1)$. We apply minimal coupling procedure when considering curved space-time.

\section{The Model}

A theoretical model of a 3D layered material with lattice coordinates $x$, $y$ and $z$, whose inter-layer distance $\Delta z$ is much greater than the electromagnetic interaction range $d$ of the inter-layer bonds ($\Delta z \gg d$) is conceptualized to represent the honeycomb layered oxide subclass described in the previous section. We firts introduce a voltage, $V$ that produces an electric field $-\vec{\nabla}_{xy}V = (E_{x}, E_{y}, 0)$ in the $x$ and $y$ direction along the $x-y$ (honeycomb) plane of the crystal material; and then preclude electromagnetic interactions and classical motion of ${\rm A}$ cations along the $z$ direction (shown in {\bf Figure 2}) which we assume are negligible due to the condition $\Delta z \gg d$, where $d$ is also the screening length for the electromagnetic field along the $z$ direction.

Considering a single honeycomb layer; viewing the cations as a fluid of charge density $j^{0}$, we can introduce the charge density vector $j^{a} = (j^{0}, \Vec{j})$ to impose the local charge conservation on the $x - y$ plane using the divergence condition $\partial_{a}j^{a} = 0$. This leads to the solution, $j^{a} = \sigma_{xy} \varepsilon^{abc}\partial_{b}A_{c}\equiv \sigma_{xy}(B_{z}, -E_{x}, E_{y})$ where $\sigma_{xy}$ is the conductivity of the cations in the $x-y$ plane, $A_{a}$ is the electromagnetic vector potential and $\varepsilon^{abc}$ is the totally anti-symmetric Levi-Civita symbol. This solution means that electromagnetic theory in 2D naturally leads us to the Chern-Simons term $\varepsilon^{abc}\partial_{b}A_{c}$.\cite{Chern_Simons1974, Dunne1999} This term contains only three electromagnetic fields: the $z$ component of the magnetic field pointing in the $z$ direction, $B_{z} = \partial A_{y}/\partial x - \partial A_{x}/\partial y$ and the $x$ and $y$ components of the electric field, $E_{x} = \partial A_{t}/\partial x - \partial A_{x}/\partial t$ and $E_{y} = \partial A_{t}/\partial y - \partial A_{z}/\partial t$ pointing in the $x$ and $y$ direction respectively, as displayed in {\bf Figure 3(a)} \& {\bf 3(b)}.

\begin{figure}
\begin{center}\label{Honeycomb}
\includegraphics[width=1\columnwidth,clip=true]{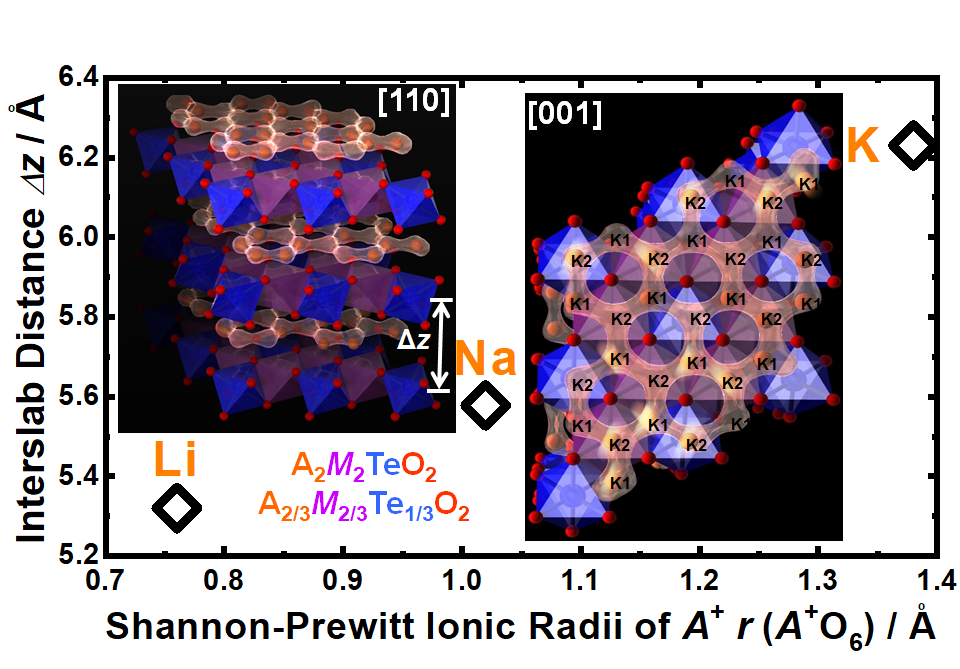}
  \caption{\textbf{Figure 2: Trend in increase of the inter-layer distance ($z$) for honeycomb layered oxides, adopting the composition ${\rm A^{+}_{2}Ni_{2}^{+2}Te^{6+}O_{6}\, (A^{+}_{2/3}Ni^{2+}_{2/3}Te^{6+}_{1/3}O_{2})}$, with increase in ionic radius of ${\rm A}$ cations.} \blue{\textbf{Inset (left)} shows a polyhedral view of the layered structure of ${\rm A_{2}Ni_{2}TeO_{6}}$ along the $z$-axis: ${\rm A}$ ions are brown spheres, ${\rm O}$ ions are small red spheres, ${\rm Ni}$ and $\rm Te$ atoms are enclosed within the purple and blue octahedra of oxygen atoms respectively. \textbf{Inset (right)} depicts a fragment of the honeycomb structure along the $x-y$ plane with a two-dimensional (2D) motion of ${\rm A}$ ions that reside above or below the honeycomb slabs.}}
\end{center}
\end{figure}

\subsection{Ansatz 1}Given that, in the absence of the applied voltage the cations form a 2D honeycomb lattice, the diffusion current (mobile cations) across the honeycomb lattice is assumed to be extracted by the potential energy of the applied voltage/electric fields (as shown in {\bf Figure 3(b)}). A correlation between the total number of these mobile cations ($g \in integer$) (as in {\bf Figure 3(c)}) and the quasi-stable 2D configurations shown in {\bf Figure 3(d), 3(e)} \& {\bf 3(f)} can be inferred, since each configuration is expected to supply a unit charge $q$ (where $q = +e \simeq 1.6 \times 10^{-19}$ C for ${\rm A = K, Na, Li, \textit{etc}}$) which leaves a vacancy in the lattice while simultaneously constituting a diffusion current equivalent to the spatial component of the Chern-Simons term,
\begin{align}\label{current_density_eq}
\vec{j} = q\rho_{xy}(\vec{n}\times\vec{v}) = \sigma_{xy}(\vec{n}\times\vec{E}),
\end{align}
where $\vec{v} = (v_{x}, v_{y}, v_{z})$ is the velocity vector, $\vec{E} = (E_{x}, E_{y}, E_{z})$ is the applied electric field, $\vec{n} = (0, 0, 1)$ is the unit vector normal to the honeycomb surface oriented in the $x-y$ plane, $\rho_{xy} = -K(x,y)/4\pi d$ is their 3D number density
, $K(x,y)$ is the Gaussian curvature\cite{Chavel1994} of the honeycomb surface $M$ after extraction of $g$ cations which satisfies the definite integral (Gauss-Bonnet theorem\cite{Allendoerfer1943}) $\int_{M}d(Area)K(x,y) = 2\pi\chi$ and $\chi = 2 - 2g$ is the Euler characteristic of the unbounded surface $M$ with $g \in integer$ the genus of $M$. This means that $\int_{M}d(Area)\,\rho_{xy} \approx g/d$ for $g \gg 1$ leading to $\int_{M} d(Vol)\,\rho_{xy}\equiv \int_{M}d(Area)\int_{d}dz\,\rho_{xy} (x,y) \approx g$. Thus, the Gauss-Bonnet theorem sets further constraints on our model by linking the Gaussian curvature and the genus of the honeycomb surface\footnote{Treated as a differentiable manifold} to transport quantities related to the electrodynamics of the cations such as their number density.

\begin{figure}
\begin{center}\label{Gauss_Bonnet_Honeycomb}
\includegraphics[width=0.75\columnwidth,clip=true]{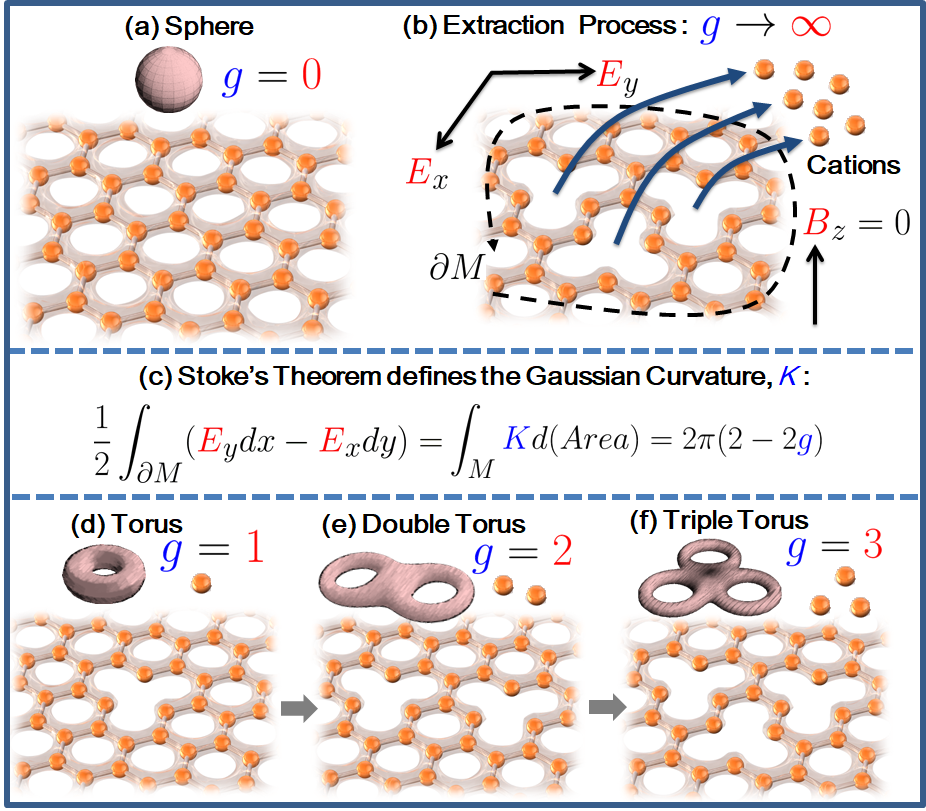}
  \caption{\textbf{Figure 3: Various quasi-stable configurations of the cation honeycomb layers and the directions of the electric and magnetic field.} \blue{\textbf{(a)} The cations arranged in a honeycomb fashion with no cationic vacancies ($g = 0$) topologically equivalent to the sphere. \textbf{(b)} The extraction process of $g \rightarrow \infty$ cations from the honeycomb surface by applied electric fields $E_{x}$ and $E_{y}$. The magnetic field in the $B_{z} = 0$ is taken to be zero. Vacancies created by this extraction process can be counted by tracking the change in the electric fields at the boundary of the honeycomb surface $\partial M$. \textbf{(c)} The equation that tracks the changes where $m$ and $q$ are the mass and charge of a single cation and $K$ the Gaussian curvature of the surface. Applying Stokes's theorem transforms the equation to the Gauss-Bonnet theorem where $g$ is the genus. \textbf{(d)} The honeycomb surface with vacancy $g = 1$ equivalent to a torus. \textbf{(e)} The honeycomb surface with vacancy $g = 2$ equivalent to a double torus. \textbf{(f)} The honeycomb surface with vacancy $g = 3$ equivalent to a triple torus.}}
\end{center}
\end{figure}

\subsection{Ansatz 2}\label{Ansatz_1}
To incorporate quantum theory along side diffusion occurring along the 2D honeycomb layer, we introduce a second ansatz,
\begin{subequations}\label{ansatz_eq}
\begin{align}\label{action_eq}
\int dt\,\expval{\frac{\partial}{\partial t}}{\psi} = iS(t,\vec{x}) + \frac{1}{2}\int d\vec{x}\cdot\vec{n}\times D^{-1}\vec{v},
\end{align}
where $S(t,\vec{x})$ is the classical action, $\ket{\psi}$ is the quantum mechanical wavefunction (kernel) of the charged fluid of the cations respectively and $D$ is the diffusion coefficient of the cations. Note that for a finite velocity $d\vec{x}/dt = \vec{v}$, the kernel has the solution $\ket{\psi} = \exp(iS)$ as expected\footnote{by Feynman's path integral reformulation of quantum mechanics\cite{Feynman2005}} (and thus satisfies the normalisation condition, $\bra{\psi}\ket{\psi} = 1$), since the second term becomes $\frac{1}{2}\int dt\, (d\vec{x}/dt)\cdot\vec{n}\times D^{-1}\vec{v} = \frac{1}{2}\int dt\,(dx^{k}/dt)D^{-1}\varepsilon_{ijk}n^{i}v^{j} = 0$ which identically vanishes in an open path. 

In contrast, the second term need not to vanish for a closed path $\partial M$ Thus, plugging eq. (\ref{current_density_eq}) into eq. (\ref{action_eq}) for a closed path yields, 
\begin{align*}
\frac{1}{2}\oint_{\partial M}d\vec{x}\cdot\vec{n}\times\frac{\sigma_{xy}}{qD\rho_{xy}}\vec{E}
= \frac{1}{2}\int_{M}d(Area)\,\vec{\nabla}_{xy}\cdot\frac{\sigma_{xy}}{qD\rho_{xy}}\vec{E} = \frac{1}{2}q\beta\int_{M}d(Area)\vec{\nabla}_{xy}\cdot\vec{E},
\end{align*}
where we have applied Stoke's theorem, Einstein-Smoluchowski relation $D = \mu \beta^{-1}$ and the Langevin result for ionic conductivity $\sigma_{xy} = q^{2}\mu\rho_{xy}$ to arrive at our result. It is now evident that non-vanishing electric fields which constitute a diffusion current require the modification of the Kernel via this second term as per ansatz 2.  For electrodynamics described by Chern-Simons theory,the 2D charge density is proportional to the flux within the boundary $\partial M$\cite{Zee2010} shown in {\bf Figure 3(b)}.\footnote{In Maxwell's theory, the charge density is given by the Gauss' law of electromagnetism $\vec{\nabla}\cdot \vec{E} = 4\pi J^{0}/\epsilon$, where $J^{0}$ is the charge density and $\epsilon$ is the permittivity of the material.} Thus, we substitute $\vec{\nabla}_{xy}\cdot\vec{E} = 8\pi\rho_{xy}/q = -2K(x,y)/qd$ \footnote{The factor of $8\pi$ is required for consistency with eq. (\ref{QG_eq})} and the argument of $\ket{\psi}$ transforms as $S \rightarrow S + i2\pi\beta m\Phi$ where we have defined the potential $\Phi$ as,
\begin{align}\label{Phi_def_eq}
    \int d(Area)\,K(x,y) \equiv -2\pi\Phi(x,y).
\end{align}
Consequently, the solution for $\ket{\psi}$ transforms to, 
\begin{align}\label{wavefunction}
\ket{\psi} = \exp i(S - i2\pi\beta m\Phi).
\end{align}
\end{subequations}
This requires that $\Phi(x,y) \equiv \frac{1}{4\pi}\int d\vec{x}\cdot\vec{n}\times\nu\vec{v} \rightarrow \vec{\nabla}_{xy}\cdot\nu\vec{v} = -2K(x,y)$ such that $1/\nu = m\mu = mD/k_{\rm B}T$ is the mean free time between collisions of the cations\footnote{The definitions $\Phi(x,y) = (4\pi m)^{-1}\int d\vec{x}\cdot\vec{n}\times\mu^{-1}\vec{v}$ and $\vec{\nabla}_{xy}\cdot\vec{E} = 8\pi\rho_{xy}/q$ are consistent with the Langevin equations $0 = d\vec{p}/dt = -\mu^{-1}\vec{v} + q\vec{E} + \vec{\eta}$ and $0 = d\vec{p}/dt = -\mu^{-1}\vec{n}\times\vec{v} + q\vec{n}\times\vec{E} + \vec{\eta}$ with the random force $\vec{\eta} = 0$. See also eq. (\ref{Langevin_eq}) and eq. (\ref{WFL2_eq}).} where $\Phi(x,y)|_{M} = -\chi$ and $m = 1/d$ acts as the effective mass of the cations. 

Notice that $\Phi(x,y)$ is reminiscent of a fictitious imaginary Aharonov-Casher phase\footnote{Aharonov-Casher phase $\gamma_{\rm AC} = -\int_{\partial M} d\vec{x}\cdot\vec{\mu}\times\vec{E}$ is the geometric phase acquired by the wavefunction of a neutral particle along a path $\partial M$ around charges, where $\vec{\mu}$ is the magnetic moment of the neutral particle.\cite{Aharonov_Casher1984, Kanyolo2019b}} and $\int d(Vol) \psi^{*}\psi \equiv \bra{\psi}\ket{\psi} = \exp(4\pi\beta m\Phi) \equiv \langle n \rangle$ takes the form of a Boltzmann factor \footnote{Alternatively, since the wavefunction is no longer normalised when $\Phi \neq 0$, it is clear that $\langle n \rangle \equiv Z$ can also be interpreted as a Lehmann weight renormalising the wavefunction $\ket{\psi} \rightarrow \ket{\Psi} = \sqrt{Z^{-1}}\ket{\psi}$ and $\bra{\Psi}\ket{\Psi} = 1$.} which gives the average number of free cations constituting the diffusion current. Setting $\psi = \langle \rho_{xy}\rangle^{1/2}\exp(iS)$ and $\ket{\psi} = \langle n\rangle^{1/2}\exp(iS)$ implies $\int d(Vol) \psi^{*}\psi =  \int d(Vol) \langle\rho_{xy}\rangle = -\langle\chi\rangle/2 = \langle g\rangle - 1$ leading to the bosonic identity,
\begin{subequations}
\begin{align}\label{Identity_eq}
    \langle g\rangle - \langle n\rangle = 1.
\end{align}
For the second quantisation formalism, the kernel can be defined as a Bogoliubov transformation,\cite{Wagner1986}
\begin{align}\label{Bogoliubov_eq}
    \hat{\psi} = \langle g \rangle^{1/2}\hat{a} + \langle n \rangle^{1/2}\exp(iS)\hat{a}^{\dagger},\\
    \hat{\psi}^{\dagger} = \langle g \rangle^{1/2}\hat{a}^{\dagger} + \langle n \rangle^{1/2}\exp(-iS)\hat{a},
\end{align}
\end{subequations}
where $\hat{a}$ and $\hat{a}^{\dagger}$ are harmonic oscillator annihilation and creation operators respectively satisfying the commutation relation $[\hat{a},\hat{a}^{\dagger}] = [\psi,\psi^{\dagger}] = 1$ which is equivalent to eq. (\ref{Identity_eq}). Thus, the  operators in eq. (\ref{Bogoliubov_eq}) act on the harmonic oscillator ground state as $\hat{a}\ket{0} = 0$ and $\bra{0}\hat{a}^{\dagger} = 0$ which guarantees that we recover the Kernel by $\hat{\psi}\ket{0} = \ket{\psi}$ and $\bra{\psi} = \bra{0}\hat{\psi}^{\dagger}$. Finally, Boltzmann's entropy formula, $\mathcal{S} = k_{\rm B}\ln g$, where the genus $g \geq 1$ is taken as the number of microstates in the system, guarantees the entropy vanishes when $n = 0$ where the $g = 1$ torus given in {\bf Figure 3(d)} represents the ground state of the system.

\subsection{Typical diffusion dynamics in 3D versus our 2D model}

We first note that typical diffusion is effectively described by Brownian motion. Consequently, the charge density function $\rho$ of diffusive charges at equilibrium temperature satisfies the 3D Fokker-Planck equation,\cite{Risken1996}
\begin{subequations}\label{Fokker_Planck_3D_eq}
\begin{align}
    0 = \frac{\partial \rho}{\partial t} = -\vec{\nabla}\cdot \rho\vec{v} + \vec{\nabla}\cdot D \vec{\nabla}\rho,
\end{align}
equivalent to the Langevin equation 
\begin{align}\label{Langevin_eq}
    0 =  m\frac{d\vec{v}}{dt} = -\frac{1}{\mu}\vec{v} + q\vec{\nabla}V,
\end{align}
\end{subequations}
where the Boltzmann factor $\rho \propto \exp-\beta qV$
is typically used to derive $\sigma_{xy} = q^2\mu\rho_{xy} = q\mu\rho$ and the Einstein-Smoluchowski relation $D/k_{\rm B}T = \mu$.

In contrast, the dynamics of the cations confined in 2D is captured by the Boltzmann factor $\langle n \rangle = \exp(4\pi\beta m\Phi)$ instead, where $4\pi m\Phi = \int d\vec{x}\cdot \vec{n}\times\mu^{-1}\vec{v} \not\equiv qV$ defined in ansatz 1 and 2, is directly linked to the topology of the 2D surface. Thus, setting $\langle n \rangle \propto \rho$, we find that the appropriate 2D Fokker-Planck equation at equilibrium for $\rho$ is given by,
\begin{align}\label{Fokker_Planck_2D_eq}
    0 = \frac{\partial \rho}{\partial t} = -\vec{\nabla}\cdot (\vec{n}\times\rho \vec{v}) + \vec{\nabla}\cdot D \vec{\nabla}\rho. 
\end{align}
The Fokker-Planck equation is solved by Fick's first law of diffusion, $\vec{j} = D\vec{\nabla}\rho$, together with the Chern-Simons current density $\vec{j} = \sigma_{xy}\varepsilon^{iab}\partial_{a}A_{b}$ to yield eq. (\ref{current_density_eq}).

\subsection{The kernel in 2D + 1 dimensions}
It is necessary to establish the equation of motion for the kernel $\ket{\psi} = \exp i(S - i2\pi\beta m\Phi)$. In standard quantum mechanics, $\ket{\psi}$ simply satisfies the Schrodinger equation, $i\partial \ket{\psi}/\partial t = \mathcal{E}\ket{\psi}$ where $-\partial S/\partial t = \mathcal{E}$ is the energy of the system. The presence of a finite screening length of electromagnetic interactions along the $z$ direction presupposes that the inter-layer crystalline structures such as ${\rm M{\rm O}_{6}}$ and ${\rm Te{\rm O}_{6}}$ octahedra shown in {\bf Figure 2} will act as tunnel barriers for the cations in the layers of the manifold $M$. This implies that, when $\Delta z \gg d$, the action (quantum phase) $S$ satisfies the dynamics of a large tunnel junction,\cite{Kanyolo2019}
\begin{subequations}\label{wavefunction_eq}
\begin{align}\label{wavefunction_eq1}
    \partial_{\mu}S = \frac{d}{2}qn^{\mu}F_{\mu\nu},
\end{align}
where $F_{\mu\nu} = \partial_{\mu}A_{\nu}-\partial_{\nu}A_{\mu}$ is the electromagnetic field tensor and $n^{\mu} = (0, \vec{n})$ is the unit normal four-vector along the $z$ direction. Then, it is straightforward to show that $\ket{\psi} = \sqrt{\langle n \rangle}\exp iS$ now obeys, $i\partial_{\mu}\ket{\psi} = qdn^{\nu}f_{\mu\nu}\ket{\psi}$ whose solution is given by,
\begin{align}\label{wavefunction_eq2}
   \ket{\psi} = \exp(-iqd\int dx^{\nu}n^{\mu}f_{\nu\mu}),
\end{align}
\end{subequations}
which corresponds to the transformation $S \rightarrow \ln\ket{\psi}$ and  $F_{\mu\nu}\rightarrow f_{\mu\nu} = \partial_{\mu}A_{\nu}-\partial_{\nu}A_{\mu} + i\frac{\beta}{d}\varepsilon_{\mu\nu}^{\,\,\,\,\,\,\sigma\rho}\partial_{\sigma}A_{\rho}$ in eq. (\ref{wavefunction_eq1}) where $f_{\mu\nu}$ is equivalent (in vector notation) to the Riemann-Silberstein\cite{Riemann_Silberstein} vector $\vec{F} = \vec{E} + i\frac{\beta}{d}\vec{B}$ and is dual, $\vec{G} = \vec{B} + i\frac{\beta}{d}\vec{E}$.\footnote{Strictly speaking, $f_{\mu\nu}$ is in the form of the (self-dual) Cabbibo-Ferrari tensor\cite{Fryberger1989} given by $\partial_{\mu}A_{\nu}-\partial_{\nu}A_{\mu} + \varepsilon_{\mu\nu}^{\,\,\,\,\,\,\sigma\rho}\partial_{\sigma}a_{\rho}$, where $a_{\mu} = i\frac{\beta}{d}A_{\mu}$ is the dual of the electromagnetic potential, $A_{\mu}$. This self duality is taken as a consequence of the proportionality of flux and charge in 2D.} 

\subsection{Connection of the  kernel, $\psi$ to Gaussian Curvature}

In Riemannian geometry, the Gaussian curvature of a bounded 2D manifold $M$ is given by, $K = \mathcal{R}/2$ where $\mathcal{R}$ is the Ricci scalar of the manifold $M$. Moreover, in ansatz 2, we assumed that $\vec{\nabla}_{xy}\cdot\vec{E} = 8\pi\rho_{xy}/q$, by assuming the proportionality (equivalence) of charge and flux in 2D.\cite{Zee2010} 
By introducing a Hermitian tensor $K_{\mu\nu} = R_{\mu\nu} - iqF_{\mu\nu}$ defined on a $\rm {3D} + 1$ Reimannian manifold,\footnote{$K_{\mu\nu}$ is defined to satisfy $[D_{\mu}, D_{\nu}]U^{\mu} = K_{\mu\nu}U^{\mu}$ with $D_{\mu} = \nabla_{\mu} - iqA_{\mu}$ the  covariant derivative and $U_{\mu} = \partial_{\mu}S$ the velocity four-vector.} it is possible to impose the condition,
\begin{align}\label{QG_eq}
    \nabla^{\nu}K_{\nu\mu} = -\frac{8\pi}{m}\psi^{*}\partial_{\mu}\psi,
\end{align}
where $R_{\mu\nu}$ is the Ricci tensor, $\nabla_{\mu}$ is the metric compatible covariant derivative and $\nabla_{\sigma}g_{\mu\nu} = 0$. Thus, it certainly follows that a new field $g_{\mu\nu}$ that describes the physics in the crystal\footnote{Can be viewed as an emergent gravitational field e.g. from entropic considerations\cite{Verlinde2011}} is introduced by this theory.\footnote{It is important to check the extent that $g_{\mu\nu}$ satisfies Einstein Field Equations.\cite{Misner_Wheeler_Thorn} Using eq. (\ref{Lorentz_eq}), eq. (\ref{QG_Real_eq}) and eq. (\ref{QG_Imaginary_eq}) together Bianchi identity, eq. (\ref{QG_eq})  can be shown to transform to $R^{\mu\nu} - \frac{1}{2}Rg^{\mu\nu} + \Lambda g^{\mu\nu} = -\kappa T^{\mu\nu}$ with $T^{\mu\nu} = m\rho_{xy}U^{\mu}U^{\nu} + \epsilon_{0}(F^{\alpha\mu}F_{\alpha}^{\,\,\nu} - \frac{1}{4}F^{\alpha\beta}F_{\alpha\beta}g^{\mu\nu})$, $\kappa = 8\pi/m^2$ and $\epsilon_{0} = \frac{q^2}{8\pi}$. In General Relativity, $\kappa = 8\pi G$ where $G$ is the gravitational constant and $\epsilon_{0}$ is the permittivity of free space. Comparing the two theories implies that our approach is equivalent to Einstein's gravitational theory with a charged particle of Planck mass $m = \sqrt{G^{-1}} = d^{-1}$ and Planck charge $q = \sqrt{8\pi\epsilon_{0}}$.} This is not entirely unexpected since the partition function is given by $Z \equiv \langle n \rangle = \exp(4\pi\beta m\Phi)$ and thus the free energy of the system $F = -\beta^{-1}\ln Z = -4\pi m\Phi$ is proportional to a gravitational potential $\Phi$. 

Substituting $S = m\int dx^{\mu}U_{\mu} \equiv -m\int d\tau$ in $\psi = \sqrt{\rho_{xy}}\exp(iS)$ where $U^{\mu}$ is the four-velocity vector of the cations and using the Bianchi identity $\nabla^{\nu}R_{\mu\nu} = \partial_{\mu}R/2$, the real part and imaginary parts of eq. (\ref{QG_eq}) become,
\begin{subequations}
\begin{align}\label{QG_Real_eq}
    \partial_{\mu} R = -\frac{8\pi}{m}\partial_{\mu}\rho_{xy},\\
    \nabla_{\nu}F^{\nu\mu} = 8\pi\rho_{xy}U^{\mu}/q,
    \label{QG_Imaginary_eq}
\end{align}
\end{subequations}
respectively. The solution for eq. (\ref{QG_Real_eq}) is given by $\delta R \equiv R - 4\Lambda = -(8\pi/m)\rho_{xy}$, where $R$ is the ${\rm 3D + 1}$ Ricci scalar and $4\Lambda$ is taken as the initial curvature of the manifold with no cations extracted from the $x -y$ plane by the electric field. As such, ansatz 1 requires that the curvature variation $\delta R \rightarrow \mathcal{R} = 2K$ be  projected to the ${\rm 2D}$ Ricci scalar $\mathcal{R}$ under certain conditions satisfied by the honeycomb layered oxide. \footnote{Since $K(x,y)$ is independent of $t$ and $z$, the limit $\mathcal{R} \rightarrow 2K(x,y)$ requires the Ricci scalar $R$ to also be independent of $t$ and $z$.} Particularly, this limit entails introducing a time-like Killing vector $\xi^{\mu} = (1, \vec{0})$ and a space-like Killing vector $n^{\mu} = (0, \vec{n})$, which impose the conditions $\xi^{\mu}\partial_{\mu}R = \partial R/\partial t = 0$ and $n^{\mu}\partial_{\mu}R = \partial R/\partial z = 0$ respectively.\footnote{A Killing vector $K^{\mu}$ satisfies\cite{Carroll2003} $\nabla_{\mu}\nabla_{\nu}K^{\mu} = R_{\mu\nu}K^{\mu}$. Using the anti-symmetry relation $\nabla_{\nu}K_{\mu} = - \nabla_{\mu}K_{\mu}$ and the Bianchi identity $\nabla_{\nu}R^{\mu\nu} = \frac{1}{2}\nabla^{\mu}R$, the divergence $\nabla^{\nu}\nabla_{\mu}\nabla_{\nu}K^{\mu} = \nabla^{\nu}(R_{\mu\nu}K^{\mu}) = \frac{1}{2}K^{\mu}\nabla_{\mu}R$ identically vanishes.}This in turn imposes energy and momentum conservation  (in the $z$ coordinate) in accordance with Noether's theorem.
Thus, computing the time component of eq. (\ref{wavefunction_eq}), we arrive at eq. (\ref{action_eq}),
\begin{multline*}
    \int dt \langle \psi|\frac{\partial}{\partial t}|\psi\rangle
    = \int dtd^{\,3}x \sqrt{-\rm{det}(g_{\mu\nu})} \psi^{*}\partial_{\nu}\psi\xi^{\nu}
    = \frac{-1}{8\pi d}\int d^{\,4}x \sqrt{-\rm{det}(g_{\mu\nu})} \nabla_{\mu}K^{\mu}_{\,\,\,\nu}\xi^{\nu} \\
    = \frac{-1}{8\pi d}\int d^{\,4}x \sqrt{-\rm{det}(g_{\mu\nu})} \nabla_{\mu}R^{\mu}_{\,\,\,\nu}\xi^{\nu} + \frac{q}{8\pi d}i\int d^{\,4}x \sqrt{-\rm{det}(g_{\mu\nu})} \nabla_{\mu}F^{\mu}_{\,\,\,\nu}\xi^{\nu}\\
    = \frac{-1}{8\pi d}\int d^{\,4}x \sqrt{-\rm{det}(g_{\mu\nu})}\xi^{\nu}\partial_{\nu}R + \frac{q}{8\pi d}i\int d^{\,4}x \partial_{\mu}(\sqrt{-\rm{det}(g_{\mu\nu})}F^{\mu}_{\,\,\,\nu})\xi^{\nu}\\
    = \frac{-1}{4\pi d}\int d^{\,3}x \sqrt{-\rm{det}(g_{\mu\nu})}\frac{\delta R}{2} + \frac{q}{8\pi d}i\int dt \int d^{\,2}x \sqrt{-\rm{det}(g_{\mu\nu})}n^{\mu}\xi^{\nu}F_{\mu\nu}\\
    = -\beta m\int d^{\,2}x \sqrt{{\rm{det}}(g_{ij}^{\rm 2D})}\frac{\mathcal{R}}{2} + \frac{q}{8\pi d}i\int dt \int d^{\,2}x \sqrt{-\rm{det}(g_{\mu\nu})}n^{\mu}\xi^{\nu}F_{\mu\nu}\\
    = -\beta m\int d^{\,2}x \sqrt{{\rm{det}}(g_{ij}^{\rm 2D})} K  + \frac{qd}{2}i\int dt\, n^{\mu}\xi^{\nu}F_{\mu\nu}
    = 2\pi\beta m\Phi + iS,
\end{multline*}
where we have set $\sqrt{-\det (g_{\mu\nu})} \equiv 4\pi \beta m\sqrt{\det(g_{ij}^{\rm 2D})}$, $\int dx^{\mu}n_{\mu} = d$, $\int d^{\,2}x\sqrt{-\rm{det}(g_{\mu\nu})} = 4\pi d^2$, $\int d^{\,2}x \sqrt{{\rm{det}}(g_{ij}^{\rm 2D})} = \int d(Area)$ and assumed that $n^{\nu}\xi^{\mu}F_{\mu\nu}$ is fairly constant over the surface given by $\int d^{\,2}x\sqrt{-\rm{det}(g_{\mu\nu})}$. This result is equivalent to eq. (\ref{ansatz_eq}a) in ansatz 2 where we have used eq. (\ref{wavefunction_eq}) in the last line. In order to preserve the Einstein-Smoluchowski relation $\mu = \beta D$ and the 2D Fokker-Planck relation eq. (\ref{Fokker_Planck_2D_eq}), the projection requires the factor of $4\pi\beta m$ between the 3D + 1 and 2D metrics. Nonetheless, this factor can be absorbed by the redefinition of $\Phi$ in eq. (\ref{Phi_def_eq}). 

\section{Results of the Model}

\subsection{Electrodynamics of the cations.}

Notice that eq. (\ref{QG_eq}) is invariant under the gauge transformation $\partial_{\mu} \rightarrow \partial_{\mu} - iqb_{\mu}$ and $S = -m\int d\tau \rightarrow -m\int d\tau + q\int b_{\mu}U^{\mu} d\tau$ where $b_{\mu}$ is a gauge potential. The choice $b_{\mu} = \frac{1}{4}\int\varepsilon_{\mu\nu}^{\,\,\,\,\,\,\sigma\rho}\partial_{\sigma}A_{\rho}dx^{\nu}$ with $\varepsilon_{\mu\nu}^{\,\,\,\,\,\,\sigma\rho}$ the Levi-Civita symbol where $\epsilon^{0i\sigma\rho}\partial_{\sigma}A_{\rho} = B_{i}$ and  $\epsilon^{ij\sigma\rho}\partial_{\sigma}A_{\rho} = \varepsilon^{ijk}E_{k}$ are the magnetic and electric fields respectively, leads to the following equation of motion for the cations.\footnote{Deriving the equation of motion for the cations: $\partial_{\mu}S = mU_{\mu} + qb_{\mu}$ and $\nabla_{\nu}\partial_{\mu}S = m\nabla_{\nu}U_{\mu} + q\nabla_{\nu}b_{\mu}$. Using the fact that $\nabla_{\nu}\partial_{\mu}S = \nabla_{\mu}\partial_{\nu}S$ is symmetric, we find $m\nabla_{\nu}U_{\mu} - m\nabla_{\mu}U_{\nu} = q\nabla_{\mu}b_{\nu} - q\nabla_{\nu}b_{\mu}$. Contracting the expression with $U^{\nu}$ and using $U^{\mu}U_{\mu} = -1 \rightarrow \nabla_{\nu}(U^{\mu}U_{\mu}) = 2U^{\mu}\nabla_{\nu}U_{\mu} = 0$, we find $mU^{\nu}\nabla_{\nu}U_{\mu} = qU^{\nu}(\nabla_{\mu}b_{\nu} - \nabla_{\nu}b_{\mu})$ equivalent to eq. (\ref{Lorentz_eq}).}
\begin{subequations}
\begin{align}\label{Lorentz_eq}
    mU^{\nu}\nabla_{\nu}U^{\mu} = \frac{q}{2}U_{\nu}\varepsilon^{\nu\mu\sigma\rho}\partial_{\sigma}A_{\rho}. 
\end{align}
Recall, we introduced a $t$-like vector $\xi^{\mu}$ and $z$-like Killing vector $n^{\mu}$ to guarantee the independence of the Ricci scalar $R$ on these coordinates. In the weak field limit\footnote{The weak field limit entails taking the $small\,corrections$ to be related to the Newtonian potential. In particular, since $U^{\mu}\nabla_{\mu}U^{\nu} = dU^{\nu}/d\tau + \Gamma^{\nu}_{\,\,\sigma\rho}U^{\sigma}U^{\rho}$, where $\Gamma^{\nu}_{\,\,\sigma\rho} = g^{\mu\nu}(\partial_{\sigma}g_{\rho\mu} + \partial_{\rho}g_{\sigma\mu} - \partial_{\mu}g_{\sigma\rho})/2$ is the Christoffel symbol, in the weak field limit given by ${\rm diag}(g_{\mu\nu}) = (g_{00}, -1, -1, -1)$, the symbol is non-vanishing only for $\Gamma^{i}_{\,\,00} = -\vec{\nabla}g_{00}/2$.\cite{Carroll1997}} ${\rm diag}(g_{\mu\nu}) \simeq (1,-1,-1,-1) + small\,corrections$ with $U^{\mu} \simeq \exp(\Phi)(1, \vec{\mathcal{V}})$ and $U_{\mu} \simeq \exp(\Phi)(1, -\vec{\mathcal{V}})$ such that $U^{\mu}U_{\mu} = -1 \rightarrow U^{0} = dt/d\tau = \exp(\Phi) = 1/\sqrt{1 - \mathcal{\mathcal{V}}^2}$, eq. (\ref{Lorentz_eq}) reduces to,
\begin{align}\label{WFL_eq}
    m\frac{dU^{0}}{dt} = \frac{q}{2}\vec{\mathcal{V}}\cdot\vec{B},\\
    \frac{d \vec{\mathcal{P}}}{dt} = mU^{0}\vec{\nabla}g_{00}/2 + \frac{q}{2}\vec{B} + \frac{q}{2}(\vec{\mathcal{V}}\times\vec{E}),
    \label{WFL2_eq}
\end{align}
where $\vec{\mathcal{P}} = mU^{0}\vec{\mathcal{V}}$ is the momentum vector, $g_{00}$ is the time-time component of the metric tensor related to $\xi^{\mu} = (1, \vec{0})$ by $g_{00} = \xi^{\mu}\xi_{\mu}$. Substituting  eq. (\ref{WFL_eq}) into eq. (\ref{WFL2_eq}) and identifying $\vec{\mathcal{V}} = \vec{n}\sqrt{1 - (1/U^{0})^2}$ as the velocity vector of the cations (pointing solely in the $z$ direction), reduces eq. (\ref{Lorentz_eq}) to, $\frac{1}{2}\vec{n}\cdot\vec{\nabla} g_{00} \equiv -\vec{n}\cdot\vec{\nabla} \Phi = 0$, $\frac{\partial}{\partial t}\Phi = \frac{q}{2m}\vec{n}\cdot\vec{B}\sqrt{(1/U^{0})^2 - (1/U^{0})^4} \simeq \frac{q}{2m}\vec{n}\cdot\vec{B}\sqrt{2\Phi}$ and $-\vec{\nabla}\Phi \equiv -\frac{1}{2}\vec{\nabla}g_{00} = \frac{q}{2m}\vec{n}\times\vec{E}\sqrt{(1/U^{0})^2 - (1/U^{0})^4} \simeq \frac{q}{2m}\vec{n}\times\vec{E}\sqrt{2\Phi}$, where we have substituted $g_{00} = 1 + 2\Phi$ thus identifying $\Phi$ as a Newtonian potential satisfying the weak field limit $0 \leq \Phi \ll 1$ and $U^{0} = \exp(\Phi) \simeq 1 + \Phi$. 
Finally, rescaling $\Phi \rightarrow (2\pi\Phi)^2/2$ allows one to write,
\begin{align}
\partial^{a}\ln \langle n \rangle = 4\pi m\partial^{a}\Phi \simeq \varepsilon^{abc}\partial_{b}A_{c},
\end{align}
\end{subequations}
which is consistent with eq. (\ref{wavefunction_eq2}).\footnote{This rescaling corresponds to appropriately rescaling the $\vec{E}$ and $\vec{B}$ fields.}

\subsection{Equilibrium properties of the cations near $g \simeq 1$}

In the weak field static limit when the magnetic field vanishes as shown in {\bf Figure 3(b)}, $0 = \partial\Phi/\partial t = qB_{z}/4\pi m$, eq. (\ref{QG_Real_eq}) reduces to,
\begin{align}\label{Liouville_eq}
    \frac{1}{2}\nabla^2g_{00} = \frac{4\pi}{m}\rho_{xy}U^{0}U^{0} \rightarrow \nabla_{xy}^2\Phi = -K\exp(2\Phi).
\end{align}
Equation (\ref{Liouville_eq}) takes the form of the well-known Liouville's equation where $K = -4\pi\rho_{xy}/m$ is evidently the Gaussian curvature. To achieve constant conductivity $\sigma_{xy} \propto \rho_{xy}$ at equilibrium, the charge density should vary spatially very slowly and an approximation can be made to solve eq. (\ref{Liouville_eq}) with a constant Gaussian curvature $\rho \propto K = K_{0} + \delta K \simeq K_{0}$. Since the shift from 3D to 2D ($\vec{\nabla} \rightarrow \vec{\nabla}_{xy}$) is a consequence of imposing the Killing vector $n^{\mu}$ on the metric tensor $g_{\mu\nu}$ in 3D + 1 dimensions, we first solve eq. (\ref{Liouville_eq}) in 3D to yield, $g_{00} = 1 - \ln K_{0}r^2 = 1 + 2\Phi$ with $\vec{x} = (x,y,z)$, $\vec{x}_{0} = (x_{0},y_{0},z_{0})$ and $\vec{x} - \vec{x}_{0} = \vec{r}$, then restrict the solution to the $x-y$ plane by setting $z = z_{0}$. 

Subsequently, this solution can be 
plugged into eq. (\ref{Phi_def_eq}) to yield $\int_{M} K(r)d(Area) = 2\pi\chi = -2\pi\Phi|_{M} = -\pi \ln(K_{0}r^2)^{-1}|_{M} = -2\pi\chi\ln(K_{0}r^2)^{-2/\chi}|_{M}$ which requires that $(K_{0}r^2_{M})^{-2/\chi} \rightarrow \exp(-1) = \lim_{\chi/2 \rightarrow 0} (1 - \chi/2)^{2/\chi}$ to satisfy the Gauss-Bonnet theorem. Thus, this approximation suggests that perturbing $K_{0}$ around the ground state, $\chi/2 = 1 - g \rightarrow 0$ of the system leads to $(K_{0}r^{2}_{M})^{-1} \rightarrow 1 - \chi/2 = g$, an expression  that characterizes the discrete areas of the manifold, $Area(g) \sim K_{0}^{-1} = gr_{M}^2$, where in our case we can freely choose $Area(g = 1) \sim r_{M}^2 \equiv \mu$  to be the unit area of the quasi-stable configuration with $g = 1$ vacancy as depicted in {\bf Figure 3(d)}. Thus, through this expression the entropy of the system
$\mathcal{S} = k_{\rm B}\ln g = -k_{\rm B}\ln(K_{0}r_{\rm M}^2)$ is derived.

Moreover, in Chern-Simons theory, the conductivity $\sigma_{xy}$ is inversely proportional to the level $g \in integer$ of the Chern-Simons action,\cite{Zee2010} $S_{\rm CS} = \int d^{\,3}x\,\frac{g}{4\pi}\varepsilon^{abc}a_{a}\partial_{b}a_{c} - \frac{q^2}{2\pi}\varepsilon^{abc}a_{a}\partial_{b}A_{c} + dj^{a}A_{a}$, which leads to an effective action $S^{\rm eff}_{\rm CS} = \ln i[\int d[a_{\mu}]\exp(iS_{\rm CS})] = \int d^{\,3}x\,\frac{q^2}{4\pi g}\varepsilon^{abc}A_{a}\partial_{b}A_{c} + dj^{a}A_{a}$. Varying the effective action with respect to $A_{a}$, we get the Chern-Simons current density $j^{a} = (-q^2/2\pi dg)\varepsilon^{abc}\partial_{b}A_{c}$. Comparing it with eq. (\ref{current_density_eq}), we find $-q^2/2\pi dg = \sigma_{xy}$. Using $\sigma_{xy} = q^2\rho_{xy}\mu$ and $K_{0} = -4\pi d\rho_{xy}$, we find $-q^2/2\pi dg = \sigma_{xy} = q^2\rho_{xy}\mu = -q^2K_{0}\mu/4\pi d = -q^2\mu/4\pi d r_{M}^2g$, which implies that the mobility of the cations is given by $\mu = 2r_{M}^2$. 

\subsection{The Bose-Einstein Condensate of Cations}

We define the order parameter of the cations in the 2D honeycomb layer as $\Psi = \sqrt{\rho_{xy}}\exp(i\theta)$ where $\theta = \int d\vec{x}\cdot\vec{n}\times D^{-1}\vec{v}$. Plugging in this definition together with $\Phi = \Phi_{0} + \delta \Phi$ and $\rho_{xy} \propto \exp(4\pi m\Phi)$ in eq. (\ref{Liouville_eq}) where $\delta \Phi \sim small$ is an infinitesimal fluctuation of $\Phi$ from its initial ground state value given by, 
\begin{subequations}
\begin{align}\label{entropy_eq}
    \Phi_{0} = -\frac{1}{2}\ln (K_{0}\mu/2) \sim \frac{1}{2}\ln g
\end{align}
we find, 
\begin{multline*}
    -\frac{4\pi\exp(2\Phi_{0})}{m}|\Psi|^2 
    = \frac{1}{4\pi\beta m}\nabla_{xy}^{2}\ln |\Psi|^2 
    = \frac{1}{4\pi\beta m}\vec{\nabla}_{xy}\cdot\left ( \frac{\vec{\nabla}_{xy}|\Psi|^2}{|\Psi|^2} \right )\\ 
    = \frac{1}{2\pi\beta m}\left (\frac{\nabla_{xy}^{2}|\Psi|}{|\Psi|} - \frac{\vec{\nabla}_{xy}|\Psi|\cdot\vec{\nabla}_{xy}|\Psi|}{|\Psi|^{2}} \right ) 
    = \frac{1}{2\pi\beta m}\left (\frac{\nabla_{xy}^{2}|\Psi|}{|\Psi|} \right ) - 8\pi\beta m(\vec{\nabla}_{xy}\Phi)^2. 
\end{multline*}
Recalling that since $\Phi = \frac{1}{4\pi}\int d\vec{x}\cdot\vec{n}\times\nu\vec{v}$, we get,
\begin{align*}
    \frac{\beta\nu m}{2}(\vec{n}\times\vec{v})\cdot(\vec{n}\times\vec{v}) - \frac{1}{2\beta\nu m}\left (\frac{\nabla_{xy}^{2}|\Psi|}{|\Psi|} \right ) - \frac{4\pi^2\exp(2\Phi_{0})}{\nu m}|\Psi|^2 = 0,
\end{align*}
which represents the sum of kinetic and potential energies. Together with local charge conservation $0 = \partial j^{0}/\partial t = -\vec{\nabla}_{xy}\cdot\vec{j}$ guaranteed by Chern-Simons theory where $\vec{j} = q\rho_{xy}(\vec{n}\times\vec{v}) = qD(\Psi^{*}\vec{\nabla}_{xy}\Psi - \Psi\vec{\nabla}_{xy}\Psi^{*})/2i$, we can re-write it as the 2D Gross-Pitaevskii equation,\cite{Gross1961, Pitaevskii1961}
\begin{align}\label{GP_eq}
    0 = i\frac{\partial}{\partial t}\Psi = \left ( -\frac{1}{2m}\vec{\nabla}_{xy}^2 - \frac{4\pi^2\beta\exp(2\Phi_{0})}{m}|\Psi|^2 \right )\Psi, 
\end{align}
where we have used $\mu = 1/m\nu = \beta D$. Equation (\ref{GP_eq}) is consistent with our treatment of the cations as bosons. Note that $\pi\beta\exp(2\Phi_{0}) \equiv a_{\rm s}$ is the cation-cation (boson-boson) scattering length where the self-interaction term is attractive. Thus, this determines the fixed potential of the perturbed ground-state as,
\begin{align}\label{entropy_eq2}
  \Phi_{0} = \frac{1}{2}\ln (a_{\rm s}k_{\rm B}T/\pi),  
\end{align}
\end{subequations}
Comparing this to eq. (\ref{entropy_eq}), we find $1/g \sim K_{0}\mu/2 = \pi/a_{\rm s}k_{\rm B}T$. Thus, the number of mobile cations increases with temperature as expected. Moreover, using the Einstein-Smoluchowski relation $\mu = D_{0}/k_{\rm B}T$ we arrive at the expression for the scattering length $a_{\rm s} = 2\pi/D_{0}K_{0}$. Note that the transition to the diffusion phase with free cations ought to occur when $g = n + 1 \geq 1$ . Plugging in $g \sim T/T_{\rm c}$, where we have defined $\pi/(a_{\rm s}k_{\rm B}) \equiv T_{\rm c}$ as the critical temperature above which a finite number of cations become diffusive, we find that $0 \leq n = (T/T_{\rm c} - 1)\Theta(T/T_{\rm c} - 1)$ with $\Theta$ the Heaviside function\footnote {$\Theta(n) = 1$ when $n \geq 0$ whereas $\Theta(n) = 0$ when $n < 0$ where $n = g - 1 = T/T_{\rm c} - 1$.} to guarantee that a transition to the conductive diffusion phase occurs at $T \geq T_{\rm c}$ corresponding to the states with $g \geq 1$ as expected.  

Finally, we  employ eq. (\ref{entropy_eq2}) to obtain the free energy of the cations using the average number $\langle n \rangle = \exp(4\pi m\Phi_{0})$. Thus, the free energy is given by $F = -\beta^{-1}\ln \langle n \rangle = -\beta^{-1}\ln \exp(4\pi m\beta\Phi_{0}) = -4\pi m\Phi_{0} \sim -2\pi m\ln (T/T_{\rm c})$. Around $T \simeq T_{\rm c}$ or equivalently $(g \simeq 1)$, the free energy becomes $F = -2\pi m\ln (1 + T/T_{\rm c} - 1) \simeq -2\pi m(T/T_{\rm c} - 1) \sim -2\pi mn$. This result can be confirmed by measuring the Arrhenius  equation $\langle n \rangle = \exp(-\beta F)$ for varied lithophile elements $\rm A = K, Na, Li, etc$ in the sub-class $\rm A_{2}^{+}L_{2}^{2+}Te^{6+}O_{6}$ using muon spectroscopic measurements where the critical temperature $T_{\rm c} \sim 1/a_{\rm s}$ is determined from the hopping rate of the cations.\cite{Mastubara2020} 

\subsection{Topological order and phase transitions with charge vortices}

Since the velocity of the mobile cations depends on the potential $\Phi = \frac{1}{4\pi m}\int d\vec{x}\cdot(\vec{n}\times\mu^{-1}\vec{v}) = -\frac{1}{2}\ln (K_{0}r^2)$, its divergence $\vec{\nabla}\Phi = \frac{1}{4\pi m}\vec{n}\times\mu^{-1}\vec{v}$ yields the velocity vector of the diffusing cations to be $m_{\rm vort.}\vec{v}_{z = z_{0}} = (\vec{n}\times\vec{r})/r^2$, which is the solution for an irrotational charge vortex satisfying $\vec{\nabla}_{xy}\times \vec{v} = 0$. The charge vortices have a mass of $m_{\rm vort.} = 1/4\pi m\mu$ and represent diffusion channels of the cations under small curvature perturbations $K - \delta K \simeq K_{0}$ around the constant conductivity $\sigma_{xy} \propto \rho_{xy}$. The kinetic energy of a single vortex can be computed as,
\begin{multline}\label{H_Vort_eq}
\mathcal{H}_{\rm vort.} = (D_{0}/2)\int_{0}^{1/m}dz\int d^{\,2}x\det(\sqrt{g_{ij}^{\rm 2D}})|\vec{\nabla}\Psi|^2\\
= (D_{0}\rho_{xy}/2m)\int d^{\,2}x\det(\sqrt{g_{ij}^{\rm 2D}}) (\vec{\nabla}\theta)^2
= (\rho_{xy}/8\pi\beta m^2D_{0})\int d^{\,2}x\det(\sqrt{-g_{\mu\nu}})\,v^2\\
= 4\pi^2\mu\rho_{xy}\int^{1/\sqrt{K_{0}}}_{r_{M}} dr/r = -2\pi^2\mu\rho_{xy}\ln(K_{0}r_{M}^2) = -2\pi m\Phi_{0}/g, \end{multline}
where we have used $\Psi = \sqrt{\rho_{xy}}\exp(i\theta)$ with $\theta = \int d\vec{x}\cdot(\vec{n}\times D_{0}^{-1}\vec{v})$, $\det(\sqrt{-g_{\mu\nu}}) = 4\pi\beta m\det(\sqrt{g_{ij}^{\rm 2D}})$, $\rho_{xy} = -mK_{0}/4\pi$, $\beta D_{0} = \mu$, $1/g = K_{0}\mu/2$ and eq. (\ref{entropy_eq}). Thus, since the emergent field $\Phi_{0} = -\frac{1}{2}\ln (K_{0}r_{M}^2)$ will vanish when the vortices in the system have no kinetic energy, $\mathcal{H}_{\rm vort.} = 0$, we conclude that the geometric description can be traced to these vortices. 

Recall that the entropy of the system is given by $\mathcal{S} = k_{\rm B}\ln g$, where $g = 1/K_{0}r_{M}^2$ is the number of microstates in the system. Note that the Boltzmann factor for a vortex-antivortex pair at equilibrium temperature $T$ for a given microstate is $\exp(-\beta g2\mathcal{H}_{\rm vort.}) = \exp(4\pi\beta m\Phi_{0}) = \langle n(g) \rangle$, which coincides with the average number of free cations. This means that the free energy of these vortices is $F \simeq -2\pi m(T/T_{\rm c} - 1)$. Thus, we discover that $T_{\rm c}$ corresponds to the Berezinskii–Kosterlitz–Thouless (BKT) transition\cite{BKT1973} temperature above which unpaired vortices appear in the system when $F < 0$. Finally, taking the sum over the microstates, we find $\sum_{g = 1}^{+\infty}\langle n(g) \rangle = \sum_{g = 0}^{+\infty}\langle n(g) \rangle - 1 = 1/(\exp(\beta 2\mathcal{H}_{\rm vort.}) - 1)$ which is simply the Bose-Einstein distribution of the vortices.

\section{Discussion}

We are left to puzzle how a geometric theory emerges in the crystal and its explicit relation to the honeycomb framework. We note that the 3D crystal is considered electrochemically neutral when the 2D honeycomb structure bears no vacancies in the lattice. However, its intrinsic Gaussian curvature given by $2\Lambda$ is not guaranteed to vanish. Since adiabatic variations from this intrinsic curvature  is modeled to occur by extraction of cations from the honeycomb surface leading to vacancies which we identify as the genus, $g$ of the surface, this links any curvature variations to changes in the number density of the extracted cations. In essence, we consider this as the origin of the geometric  description. Moreover, the physics of these curvature deformations which affect the transport properties of the cations are captured by an emergent field $\Phi$ that has a gravitational description given by Liouville's equation.

The extracted cations contribute to a diffusion current density via a Chern-Simons term that naturally arises in 2D electrodynamics. This extraction process occurs at a critical temperature $T_{\rm c} \sim 1/a_{\rm s}$ related to the cation-cation scattering length $a_{\rm s}$. In fact, we have showed that the cations, under the influence of the field $\Phi$ satisfying Liouville's equation, admits a Bose-Einstein condensate description given by eq. (\ref{GP_eq}) where curvature/charge density perturbations around the ground state of the system ($g = 1$) leads to diffusion channels satisfying the irrotational vortex condition, $\vec{\nabla}_{xy}\times\vec{v} = 0$ and can be modeled by an order parameter $|\Psi|^2 \propto n \sim (T/T_{\rm c} - 1)\Theta(T/T_{\rm c} - 1)$ representing a BKT phase transition for $T \geq T_{c} \sim 1/a_{\rm s}$. Above $T_{\rm c}$, the kinetic energy of the charge vortices leads to the emergence of the field, $\Phi$, responsible for the diffusion dynamics of the cations in 2D. This places these vortices as central to understanding the link between cationic diffusion and our emergent geometric description. 

\section{Conclusion}

An idealised model of the electrodynamics of the alkali metals cations in honeycomb layered oxide frameworks has been proposed. The model links the ionic transport of the cations to the geometry and topology of these materials by applying well-established approaches of Brownian motion, Liouville field theory and Chern-Simons theory. The central result is that ionic vacancies of the 2D honeycomb lattice are related to the Euler characteristic of the surface where the Gauss-Bonnet theorem is the charge density formula of the surface, as summarised in \textbf{Figure 3}.
Such a rich description of the transport phenomena within honeycomb layered oxide frameworks stems from the interdisciplinary approach of our model, which adopts established concepts from applied mathematics, chemistry and physics. Thus, the results presented herein elucidate previously unreported interconnections between geometry, thermodynamics and quantum theory, essential in unraveling the connection between topological order and phase transitions in these materials. Further connections, particularly the explicit relation of emergent phenomena such as charge vortices and the gravitational potential within our model remains vastly unexplored. Nonetheless, we believe the results presented herein bear particular significance across a diversity of disciplines where layered materials 
play a pivotal role in frontier applied fields of high temperature superconductivity, quantum computing and energy storage.




\begin{addendum}
 
\item[Acknowledgements]
Part of this work was conducted under the auspices of the National Institute of Advanced Industrial Science
Technology (AIST), Japan Society for the Promotion of Science (JSPS KAKENHI Grant Numbers 19K15685) and Japan Prize Foundation. We would also like to acknowledge the support in proofreading done by Edfluent.
 
\item[Competing Interests] 
The authors declare that they have no competing financial interests.
 
\item[Correspondence] 
Correspondence and requests for materials should be addressed to Titus Masese~(Email: titus.masese@aist.go.jp) and requests for clarification of aspects related to the model to Godwill Mbiti Kanyolo (Email: gmkanyolo@gmail.com)

\end{addendum}

\newpage

\section*{Appendix}

\appendix

\section{Inter-distance Tuning}

The correlation between the ionic radii of A and the inter-layer distance ($\Delta z$), as shown in \textbf{Figure 2}, presents new avenues for tuning the inter-layer and intra-layer electromagnetic couplings in order to optimize the dimensionality of the magnetic lattice. This can be achieved through the introduction of ${\rm A}$ cations with larger ionic radii that leads to  the increase of the interlayer distance. Moving from ${\rm Li, Na}$ to ${\rm K}$, it is clear that there is a propensity of the inter-layer distance to increase; which consequently should affect the transport properties (id est, diffusion nature of alkali ${\rm A}$ atoms) along the two-dimensional (2D) surface. Honeycomb layered compositions of ${\rm A^{+}_{2}L_{2}^{2+}{\rm Te}^{6+}{\rm O}_{6}}$ ${\rm (A^{+}_{2/3}L_{2/3}^{2+}{\rm Te}_{1/3}^{6+}{\rm O}_{2})}$ had initially been relegated to ${\rm A = Li, Na}$; however, recent reports have shown that such compositions can be extended to lithophile large-atoms such as K.\cite{Masese2018} This is an additional advantage as these series of honeycomb compositions may be expanded to new compositions where ${\rm A = Rb^{+}, Cs^{+}, Ag^{+}, Au^{+}, Cu^{+}, H^{+}}$, and so forth. \textbf{Figure 2} (inset, right) depicts the honeycomb arrangement of ${\rm K^{+}}$ in ${\rm K_{2/3}Ni_{1/3}Te_{1/3}O_{2}}$ ${\rm (K_{2}Ni_{2}TeO_{6})}$ ,\cite{Masese2018} which is greatly influenced by the arrangement of ${\rm NiO_{6}}$ and ${\rm TeO_{6}}$ octahedra that reside in the slabs.

\section{Quasi-stable configurations as the analogues of Tori in 2D}

We shall name each configuration as {\it (3) three-(leaf) clover} where each honeycomb area is taken as a single leaf in the clover. Consequently, each vacancy $g$ lies within a 3-leaf clover at the leaf stalk. We label each cation in the honeycomb lattice as a {\it vertex} $V$, the line connecting to their adjacent neighbour as an {\it edge} $E$ and each honeycomb unit {\it area} as $A$. Our theory then requires the number of honeycomb unit {\it areas} to be left unchanged by the extraction process.\footnote{The number of honeycomb {\it faces} $F$ does not remain unchanged.} Thus, for a honeycomb lattice with no vacancies, $A = F$ and $\chi = A - E + V = 2$ where $F$ is the number of faces. On the other hand, introducing $-g$ vacancies (equivalent to extracting $g$ cations from the honeycomb surface) each at the stalk of the 3-leaf configuration, we find $A \rightarrow A$ with ($A \neq F$), $E \rightarrow E + 3g$ and $V \rightarrow V + g$. Thus, the Euler characteristic defined above transforms to $\chi \rightarrow \chi' = A - (E + g) + (V + g) = (A - E + V) - 2g = 2 - 2g$ as expected.  
\end{document}